\documentclass{sigkddExp}

\usepackage{graphicx}
\usepackage{enumerate}
\usepackage{comment}
\usepackage{url}
\usepackage{xspace}
\usepackage{float}
\usepackage{enumitem}
\usepackage{tikz}
\usepackage{caption}
\usepackage{array}
\usepackage{fancyhdr}
\usepackage{booktabs}
\usepackage{multirow}
\usepackage{rotating}
\usepackage{makecell}
\usepackage{color}
\usepackage{colortbl}
\usepackage{bigdelim}
\usepackage{biblatex}
\newlist{questions}{enumerate}{2}
\setlist[questions,1]{label=\textbf{RQ\arabic*.},ref=\textbf{RQ\arabic*}}
\setlist[questions,2]{label=(\alph*),ref=\thequestionsi(\alph*)}
\newtheorem{definition}{Definition}
\newcommand*\emptycirc[1][1ex]{\tikz\draw (0,0) circle (#1);} 
\newcommand*\halfcirc[1][1ex]{%
  \begin{tikzpicture}
  \draw[fill] (0,0)-- (90:#1) arc (90:270:#1) -- cycle ;
  \draw (0,0) circle (#1);
  \end{tikzpicture}}
\newcommand*\fullcirc[1][1ex]{\tikz\filldraw (0,0) circle (#1);}

\begin{document}
%

\title{Blockchain for Large Language Model Security and Safety: A Holistic Survey}
%

\numberofauthors{5}
%


\author{
\alignauthor Caleb Geren\thanks{These authors contributed equally to this work.} \\
   \affaddr{Lehigh University} \\
    \email{cdg225@lehigh.edu}
\alignauthor Amanda Board$^*$ \\
    \affaddr{University of Idaho} \\
    \email{boar9227@
    vandals.uidaho.edu} \\ 
 \alignauthor Gaby G. Dagher\\
    \affaddr{Boise State University} \\
    \email{gabydagher@boisestate.edu} \\
\and
 \alignauthor Tim Andersen \\
    \affaddr{Boise State University} \\
    \email{tandersen@boisestate.edu} \\  
\and
 \alignauthor{Jun Zhuang} \\
    \affaddr{Boise State University} \\
    \email{junzhuang@boisestate.edu} \\
}

\maketitle
\begin{abstract}
With the growing development and deployment of large language models (LLMs) in both industrial and academic fields, their security and safety concerns have become increasingly critical. However, recent studies indicate that LLMs face numerous vulnerabilities, including data poisoning, prompt injections, and unauthorized data exposure, which conventional methods have struggled to address fully. In parallel, blockchain technology, known for its data immutability and decentralized structure, offers a promising foundation for safeguarding LLMs. In this survey, we aim to comprehensively assess how to leverage blockchain technology to enhance LLMs' security and safety. Besides, we propose a new taxonomy of blockchain for large language models (BC4LLMs) to systematically categorize related works in this emerging field. Our analysis includes novel frameworks and definitions to delineate security and safety in the context of BC4LLMs, highlighting potential research directions and challenges at this intersection. Through this study, we aim to stimulate targeted advancements in blockchain-integrated LLM security.
\end{abstract}

\section{Introduction}
\vspace{0.5cm}
The widespread application of large language models (LLMs) has progressed at an unprecedented pace and scale in our daily lives~\cite{teubner_welcome_2023}. Such a widespread application exposes several vulnerabilities inherent to LLMs, such as data poisoning~\cite{keshk_privacy-preserving-framework-based_2020, gong_dynamic_2023}, prompt injections~\cite{mbula_mboma_assessing_2023, shen_large_2023}, and hallucinations~\cite{kasneci_chatgpt_2023, zhang_sirens_2023, pan_privacy_2020, bender_dangers_2021}. For example, prompt injections can exploit a model's propensity to disclose information, resulting in significant data leakage, like leaking personally identifiable information (PII) to an unauthorized user~\cite{winograd_loose-lipped_2022, carlini_extracting_2023}. 
Although numerous studies have attempted to address these issues~\cite{yao_survey_2024, abdali_securing_2024}, there remains no effective mitigation strategies capable of addressing growing concerns about these issues in LLMs~\cite{winograd_loose-lipped_2022, harrer_attention_2023, lukas_analyzing_2023}. Typically, defensive strategies against these threats are implemented through established machine-learning methods, such as applying differential privacy (DP) techniques to the entire dataset to enhance privacy protections~\cite{abadi_deep-learning-dp, yan_protecting_2024}. While DP strategies are one of the crucial applications, these strategies cannot fully guarantee data privacy in LLMs~\cite{lukas_analyzing_2023} due to DP's ability to protect primarily ``by whom" data is contributed, rather than ``about whom" the data is focused on. Additionally, another common attempt to tackle the data privacy problem in LLMs is to apply federated learning (FL) techniques in the training process by distributing model training across multiple nodes to create a decentralized environment~\cite{mcmahan_communication-efficient_2017}. Naturally, this lends itself to further obscuring sensitive information in a model's corpus. However, it has been shown that by taking model weights or gradients, original data from the model can still be reconstructed~\cite{liu_federated_2021}. What's worse, federated-learning approaches are susceptible to many of the same types of attacks as large language models, such as single-point-of-failure attacks or man-in-the-middle attacks~\cite{qu_blockchain-enabled_2023}. This trend of typical approaches failing to exhaustively defend against the range of attacks now affecting LLMs continues across multiple traditional threat/defense models~\cite{shayegani_survey_2023, zhang_defending_2024, yan_protecting_2024}.

To address the limitations of the above-mentioned methods and further enhance data privacy, blockchain technology has emerged as a promising solution~\cite{mingxiao_review_2017}. It ensures data integrity through various tamper-evident mechanisms, introduces a high level of confidentiality to otherwise centralized systems, and guarantees data provenance by enabling traceable and auditable records~\cite{deshpande_distributed_nodate, cao_blockchain_2023, ali_comparative_2021}. These benefits can significantly strengthen the robustness of large language models. Integrating blockchain technology lays the foundation for stronger privacy protection, enhanced inference validation, defenses against adversarial attacks, and other security measures to be incorporated into the design of large language models. This overlapping research direction is still in its early stages. To facilitate a deeper understanding of the current landscape for emerging researchers, we conduct a comprehensive literature review in this paper to explore how blockchain technology can better serve large language models (BC4LLMs). Overall, our objectives in this survey paper are to address the following four research questions:
\begin{questions}
    \item What are the pressing LLM-related security concerns that may be addressed with blockchain technology?
    \item How can we meaningfully differentiate between security and safety in the context of BC4LLMs?
    \item In what ways can blockchain technology be used to enhance the safety of LLMs?
    \item What are prominent gaps within the BC4LLMs area, how can these gaps influence research directions, and what resources can we provide to enable potential new directions?
\end{questions}

Notably, we focus specifically on how blockchain systems may impact large language models' \textit{security} and \textit{safety}. By narrowing our scope to these dimensions, we aim to provide a more detailed analysis and categorization of seemingly disparate works, thereby encouraging targeted research advances in specific directions. In general, attacks on LLMs typically manifest in two primary ways: as direct exploitations by malicious third parties that capitalize on system vulnerabilities (security)~\cite{anderson_is_2024, xue_badrag_2024, zhang_defending_2024, hu_membership_2022, liu_prompt_2024} and as inherent risks embedded within LLM structures that expose users to potential harm without external malicious influence (safety)~\cite{fraga-lamas_fake_2020, wan_kelly_2023, seneviratne_blockchain_2022, yazdinejad_making_2020}. We base our analysis of the blockchain for LLMs (BC4LLMs) on this critical distinction between security and safety, a distinction that we underscore through explicit definitions contextualized for LLMs. To the best of our knowledge, this is the first study to rigorously define these terms in the BC4LLMs context, providing a foundation for subsequent work in this area. Furthermore, we contribute to the discourse on privacy in LLMs by delineating active and passive privacy efforts, modeled after a survey about data privacy~\cite{yan_protecting_2024}.

To distinguish our analysis of BC4LLMs from other similar works through the lenses of safety and security, {\bf we compare several reviews} in this domain. He et al.~\cite{he_large_2024} examine the relationship between LLMs and blockchain in analyzing how LLMs can further enhance blockchain systems. Mboma et al.~\cite{mbula_mboma_assessing_2023} provide an exploratory review of general integrations between blockchain and large language models, which is similar to Heston's analysis of integrating the two technologies in telemedicine~\cite{f_heston_perspective_2024}. Additionally, Salah et al.~\cite{salah_blockchain_2019}, Bhumichai et al.~\cite{bhumichai_convergence_2024}, and Dinh et al.~\cite{dinh_ai_2018} provide an overview of potential and existing technologies between blockchain and artificial intelligence in general. In short, current reviews that specifically address blockchain and LLMs lack a clear focus on the specific applications of these technologies, whereas broader reviews that encompass blockchain and AI sacrifice the depth of analysis. To close this gap, we present an overview of related surveys in Table~\ref{tab:rel_works_table}, which juxtaposes the above papers' contents with our specific focuses, highlighting the distinction in our study.

\begin{table*}[t]
\caption{\textbf{Overview of Existing Related Surveys.} We compare related surveys about blockchain techniques and LLMs from various perspectives, such as background, threat model, definition, security, safety, etc. In particular, we are interested in investigating whether (i) relevant subjects are discussed in the background section, (ii) a model of threat categorization is introduced, (iii) definitions of security and safety are proposed, (iv) security and/or safety with regards to BC4LLMs is explored, (v) future BC4LLMs work is probed, (vi) the survey focuses on LLMs for Blockchain. We denote \fullcirc, \halfcirc, and \emptycirc \ as a full, partial, and no discussion of the corresponding items.}
\label{tab:rel_works_table}
\centering
\resizebox{\textwidth}{!}{%
\begin{tabular}{@{}lccccccccc@{}}
    \toprule
    \textbf{Source} & \thead{\textit{LLMs and BC} \\ \textit{Background}} & \textit{Threat Model} & \textit{Definitions} & \multicolumn{3}{c}{\textit{Security in BC4LLMs}} & \thead{\textit{Safety in } \\ \textit{BC4LLMs}} & \thead{\textit{Future} \\ \textit{Work}} & \thead{\textit{LLMs for} \\ \textit{Blockchain}} \\
    \cmidrule(lr){5-7}
     & & & & \textit{LLM} & \textit{AI} & \textit{Non-AI} & & & \\
    \midrule
    Luo, et al. 2023~\cite{luo_bc4llm_2023} & \fullcirc & \halfcirc & \emptycirc & \halfcirc & \halfcirc & \halfcirc & \halfcirc & \fullcirc & \emptycirc \\
    \midrule
    Mboma, et al. 2023~\cite{mboma_integrating_2024} & \fullcirc & \emptycirc & \emptycirc & \halfcirc & \emptycirc & \halfcirc & \halfcirc & \emptycirc & \fullcirc \\
    \midrule
    He, et al. 2024~\cite{he_large_2024} & \fullcirc & \emptycirc & \emptycirc & \emptycirc & \emptycirc & \emptycirc & \emptycirc & \emptycirc & \fullcirc \\
    \midrule
    Heston 2024~\cite{f_heston_perspective_2024} & \halfcirc & \emptycirc & \emptycirc & \fullcirc & \halfcirc & \emptycirc & \halfcirc & \halfcirc & \halfcirc \\
    \midrule
    Salah, et al. 2019~\cite{salah_blockchain_2019} & \halfcirc & \emptycirc & \emptycirc & \emptycirc & \fullcirc & \fullcirc & \halfcirc & \emptycirc & \emptycirc \\
    \midrule
    Bhumichai, et al. 2024~\cite{bhumichai_convergence_2024} & \halfcirc & \emptycirc & \emptycirc & \emptycirc & \fullcirc & \halfcirc & \halfcirc & \emptycirc & \emptycirc \\
    \midrule
    Dinh and Thai 2018~\cite{dinh_ai_2018} & \halfcirc & \emptycirc & \emptycirc & \emptycirc & \fullcirc & \emptycirc & \halfcirc & \emptycirc & \halfcirc \\
    \bottomrule
\end{tabular}%
}
\end{table*}

We outline our main {\bf contributions} and highlight the impact to answer our research questions as follows:
\begin{enumerate}
    \item In this work, we first contribute a series of frameworks, definitions, and compiled resources. Most prominently, we propose a new taxonomy about applying blockchain techniques for LLMs in Figure~\ref{fig:Tax}. Through the proposed taxonomy, we aim to succinctly explain the relevant interactions between the blockchain techniques and corresponding LLMs' vulnerabilities [\textbf{RQ1}][\textbf{RQ3}]. To further contextualize this taxonomy and ground our discussion of existing literature, we propose two foundational definitions of safety and security specific to LLMs [\textbf{RQ2}]. Moreover, we provide a collection of datasets relevant to BC4LLMs, equipping future researchers with resources to build on the connections delineated by our taxonomy and informed by our definitions [\textbf{RQ4}].
    \item We also highlight additional components of our paper that, while supporting our main contributions, serve as valuable artifacts in the BC4LLMs space. One such artifact is our definition of specific areas within the broader concept of safety, further detailed in Table~\ref{tab:def_terms_table} [\textbf{RQ2}][\textbf{R3}]. These definitions reinforce our conceptualization of safety for LLMs. To enhance our definitions of both safety and security, we specifically address privacy within the security context, reaffirming two terms introduced by Yan et al.~\cite{yan_protecting_2024}: passive and active privacy [\textbf{RQ1}][\textbf{RQ3}]. Besides, our contextualization of LLMs within various AI sub-fields [\textbf{RQ1}] and our concise taxonomic overview of blockchain components [\textbf{RQ1}][\textbf{RQ3}] hold intrinsic value as distinct contributions to the field.
    \item Last, we conduct a comprehensive literature review in Section~\ref{sec:related_works_bc4llm}, where we classify research works across several interrelated domains, offer novel insights within these categorized domains, and align all BC4LLMs research projects with LLMs' safety and security. By conducting this review, we provide an informative perspective on the potential of utilizing blockchain techniques to enhance LLMs [\textbf{RQ1}][\textbf{RQ2}][\textbf{RQ3}][\textbf{RQ4}].  
\end{enumerate}

The remaining sections are organized as follows: In Section~\ref{sec:background}, we introduce the background of blockchain technology and large language models. In Section~\ref{sec:methods}, we describe our methodology, including the criteria used to filter works for this review and the relevant definitions that guide our analysis of the current literature. We also share our model of threat categorization, which aligns with the categorization proposed by Yao et al.~\cite{yao_survey_2024}. In Section~\ref{sec:related_works_bc4llm}, we conduct a comprehensive literature review of BC4LLMs in safety and security, examining key works in relation to our proposed taxonomy. In Section~\ref{sec:datasets}, we present datasets relevant to BC4LLMs. In Section~\ref{sec:challenges}, we address key challenges at the intersection of blockchain technology and LLMs that hinder advancement in this area. In Section~\ref{sec:future_directions}, we discuss future research directions within the field of BC4LLMs. Finally, in Section~\ref{sec:conclusion}, we summarize our efforts, providing a holistic view of the current progress of BC4LLMs.

\section{Background}
\label{sec:background}
\vspace{0.5cm}
In this section, we present an overview of blockchain as a distributed ledger technology and relate the abilities of large language models to their capacities as agents with respect to their nature as both AI models and their tendency to interact with vast quantities of data. 

\subsection{Blockchain}
\begin{figure}[h]
    \centering
    \includegraphics[width=1\linewidth]{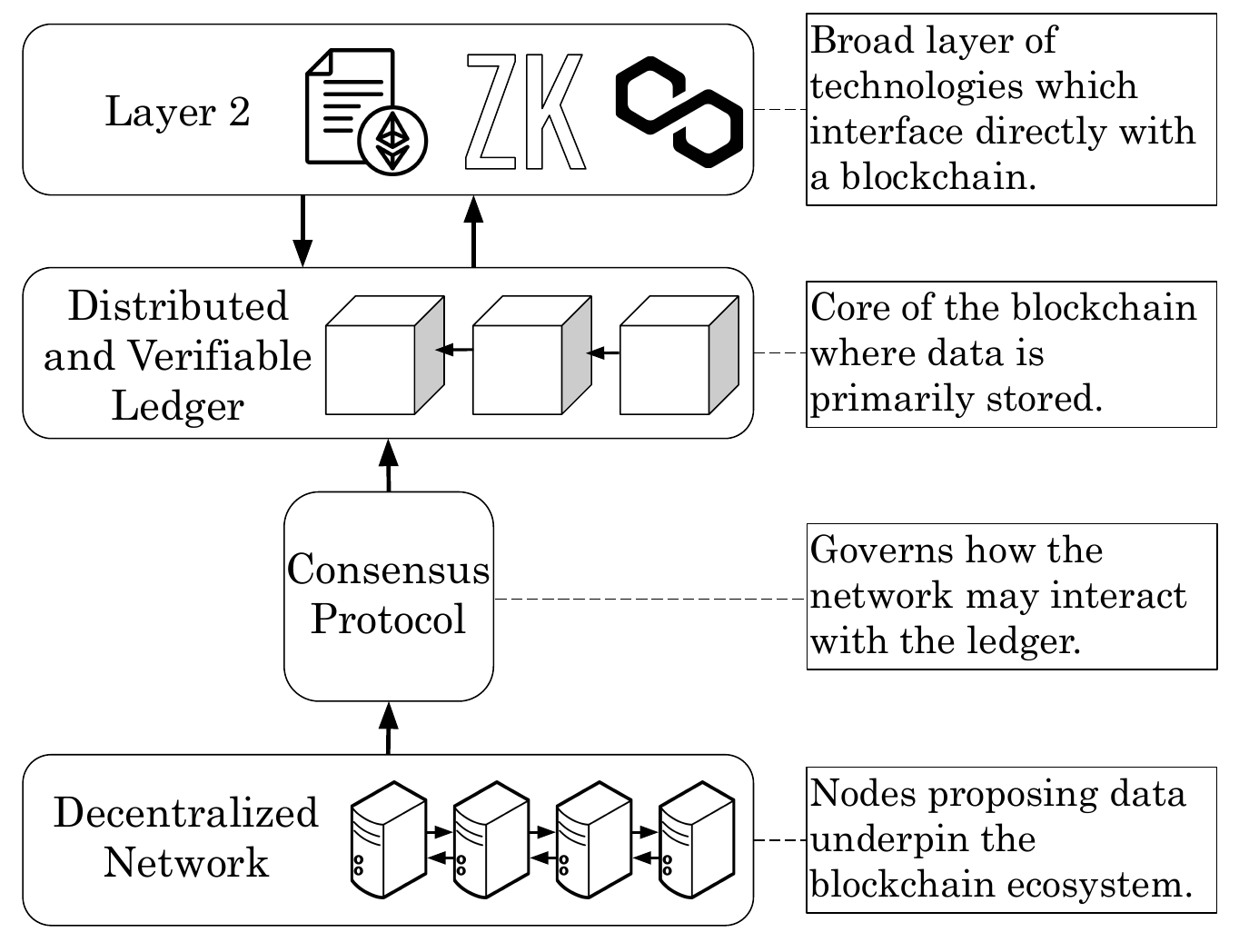}
    \caption{A blockchain consists of four main components. A decentralized network of nodes interacts with a ledger via a governing consensus mechanism. This ledger, adequately protected by the consensus mechanism, creates what we refer to as the blockchain. Layer 2 solutions can interface with this ledger to enable greater functionality between users and a blockchain's data.}
    \label{fig:blockchain_components}
\end{figure}

\vspace{0.5cm}
Since Satoshi Nakamoto~\cite{nakamoto_bitcoin_2008} introduced Bitcoin as a decentralized currency in 2008, there has been a subsequent explosion of academic and commercial interest in its underlying blockchain technology~\cite{liang2021omnilytics, song2024unveiling, song2024advancing}. Additionally, and as highlighted by the introduction of Vitalik Buterin's Ethereum blockchain in 2014~\cite{buterin_ethereum_2014}, there has been a particular focus on blockchain's potential applications in fields entirely disparate from digital currencies. The interest in blockchain, or distributed ledger technologies, stems from its guarantees of data sovereignty, transparency, and relative permanence. Concisely, these properties are often referred to as immutability and irrefutability. Ranging from many diverse fields such as health care record management, digital identity management, or tax auditing, these properties are widely applicable and highly desirable, even though the mechanisms through which we achieve them can be somewhat complex and opaque. In light of the oftentimes convoluted nature of blockchain systems, we introduce blockchain to the reader in a piecemeal fashion in order to emphasize the modular, yet interconnected nature of such systems. Figure~\ref{fig:blockchain_components} represents an overview of our characterization of blockchain systems in general. We purposefully exclude certain components such as the incentive mechanism, or wallets, as they are beyond the scope of our analysis of blockchain as a means to serve large language models. 

\subsubsection{Blockchain Components} \: \\
\textbf{Consensus Protocol.} Of particular interest to BC4LLMs, and arguably the most fundamental component within a blockchain, the consensus protocol is the governing system that controls how data is added to a blockchain's ledger. At its core is the consensus mechanism, which both ensures the validity of proposed data and fosters an environment of accountability, so that nodes submitting invalid information may be penalized accordingly. For example, the Proof of Work (PoW) consensus mechanism~\cite{nakamoto_bitcoin_2008} is by far the most widely known. In it, nodes must solve a complex mathematical equation in order to gain rights to propose data for the blockchain. When such a node submits new data, it is scrutinized by every other node in the system. If the data is malicious, or untruthful, the proposal is rejected and the corresponding processing power performed by the malicious node has effectively been wasted, as that node will not receive the incentive, a Bitcoin reward. The underlying ideas of accountability, certain nodes being selected as 'block proposers', and the 'proof' of the ability to submit information to the chain are central ideas in consensus protocols across blockchains with different consensus protocols~\cite{nguyen_proof-of-stake_2019}.

\textbf{Verifiable Ledger.} At a blockchain's core sits the verifiable ledger, a repository of data bolstered by a secure way of maintaining the integrity of that data. Of note is the particular technique through which data itself is verified on the ledger: the Merkle tree~\cite{merkle_digital_1988}, or a variation thereof. Typically implemented as a ground-up binary tree, data is stored in leaf nodes, with hashed pointers of that data cascading up the tree. This structure results in a comprehensive `Merkle root', a hash pointer consisting of all the other hash pointers in lower levels of the tree, which is ultimately based on the data stored in the leaf nodes. This technique ensures the integrity of information in the leaf nodes, as any alteration to the data is instantly reflected in the Merkle root. Likewise, new additions to the Merkle tree can be checked against previous states of the tree via a recalculation of the Merkle root accounting for the new transactions. This technique, complementing the verifiable ledger, is often the key to LLM data provenance and traceability solutions that rely on blockchain technology.

\textbf{Decentralized Network.} Critically, blockchains are decentralized networks. That is, no central server or group of servers may assume control of the network in a way that would compromise the network's state of trustlessness. This is achieved through multiple avenues, such as the aforementioned consensus protocol, the distribution of the verifiable ledger among a large number of independent nodes, and the accessibility of a given blockchain's network.~\cite{deshpande_distributed_nodate} In this way, no users in the network are required to trust any other user. This fundamental aspect of blockchain is responsible for already realized and potential advancements with LLMs concerning areas such as RAG, the training process, and even supply chain issues.

\textbf{Layer 2 Technologies.} Apart from the fundamental components found within all blockchains, several external architectures interface with blockchains and further enhance their applicability. Typically, these external architectures are referred to as layer 2 technologies, as they sit a `layer' above the `layer 1' blockchain. Increasingly relevant as blockchain's influence grows, layer 2 solutions are a burgeoning area with numerous novel research directions. Most prominent among these are smart contracts, scripts that rely on a blockchain's security guarantees to facilitate off-chain transactions~\cite{zheng_overview_2020}. Also, in layer 2, zero-knowledge rollups are often combined with the efficacy of smart contracts. Often used to strengthen scalability, zero-knowledge rollups batch unproposed transactions together, and instead of submitting the transactions themselves, submit proof that the transactions are indeed valid~\cite{sun_survey_2021}. This allows for transactions to be added on-chain without the need for every full node to redo the calculations found within those transactions. This area of layer 2 technologies is pivotal as it relates to BC4LLMs - layer 2 has the necessary dynamism to react quickly to new and emerging LLM vulnerabilities.

\subsection{Large Language Models}
\vspace{0.5cm}
In recent years, large language models (LLMs) have emerged as a pivotal force in artificial intelligence (AI), contributing to widespread applications across diverse fields, such as trustworthiness~\cite{deng2024deconstructing, jiang2024robustkv, jin2024jailbreakzoo, liu2024buckle}, scholarly document processing~\cite{zhuang2024understanding}, signal processing~\cite{ruan2024s2e}, quantum computing~\cite{liang2023unleashing}, climate production~\cite{lin2023comprehensive, lin2023mmst}, software engineering~\cite{zheng_survey_2023}, and healthcare~\cite{he_large_2024} among multiple other learning environments. Zhao et al.~\cite{zhao_survey_2023} and Yang et al.~\cite{yang_harnessing_2023} define LLMs and pre-trained language models (PLMs) from the perspectives of model size and training approach. Generally speaking, PLMs refer to language models that are pre-trained on large amounts of general text data and then fine-tuned for specific tasks. LLMs are a kind of PLM. The key distinction is that LLMs are generally larger in scale with more parameters. These large language models have demonstrated the ability to learn universal representations of language, used in various natural language processing (NLP) tasks~\cite{huang_survey_2024}, bolstering their applicability.

\begin{figure}[h]
    \centering
    \includegraphics[width=1\linewidth]{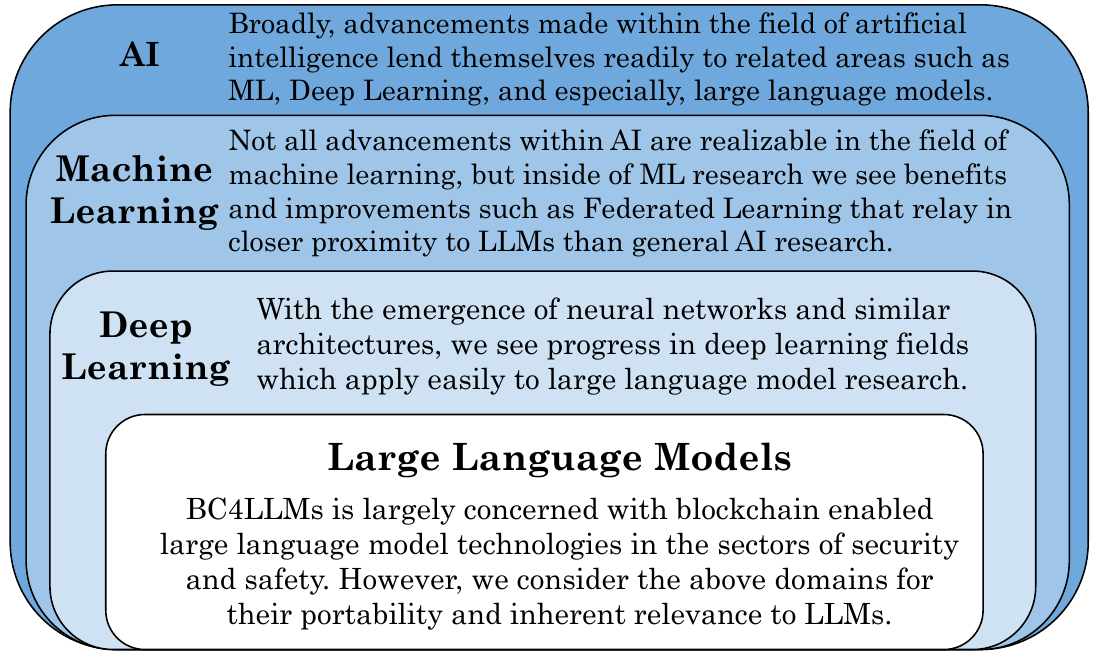}
    \caption{The connection among AI, Machine Learning (ML), Deep Learning (DL), and LLMs.}
    \label{fig:LLMandAI}
\end{figure}

We discuss connections and following developments between AI, machine learning (ML), deep learning (DL), and LLMs in Figure~\ref{fig:LLMandAI}. AI refers to a broad technique that aims at simulating human intelligence, encompassing a variety of approaches and methods. Machine Learning (ML) is a subfield of AI that develops algorithms and statistical models to automatically learn from data and efficiently perform specific tasks without the use of explicit instructions~\cite{liu_when_2021}. Deep Learning (DL) is a subset of ML that utilizes multi-layered neural networks to learn latent representation on various tasks~\cite{shafay_blockchain_2023}. LLMs are one of the popular applications using cutting-edge DL models, advancing natural language understanding and generation at the human level. In the following subsections, we elaborate on the process of how LLMs are pre-trained and utilized as safe and powerful AI systems.

\subsubsection{Model Training}
\vspace{0.5cm}
During the pre-training phase, the LLM is trained on a diverse, large dataset of textual data from various sources to learn the statistical properties of language. The LLM is equipped with a myriad of adjustable parameters, commonly reaching more than ten billion~\cite{huang_survey_2024}. Due to the huge model size and the vast amount of data used to train it, it is computationally challenging to successfully train a capable LLM, requiring distributed training algorithms for learning the model parameters~\cite{zhao_survey_2023}. Another crucial factor for LLM training is the data itself. Data that models are trained on come from a wide variety of sources, but the data itself may not be up to date~\cite{south_secure_2023}. To mitigate this shortcoming, recent advancements have introduced Retrieval Augmented Generation (RAG), which is designed to augment and rectify the information returned by LLMs by consulting up-to-date online sources. The data that the LLM was trained on also has other deficiencies, like knowledge gaps in healthcare fields where data is private and restricted~\cite{ji_survey_2023}. Due to these knowledge gaps, the LLM may conjure up hallucinations where the model generates false information during prompting~\cite{maynez_faithfulness_2020, openai_gpt-4_2024} because of a lack of relevant information. However, hallucinations may also occur with a plethora of data available as they are inherent problems in LLMs. Methods of preventing these hallucinations are elaborated in Section~\ref{BC4Misinformation}. RAG can help rectify hallucinations, and fill in the gaps of data the LLM is missing, by using up-to-date and validated information from trustworthy online resources. This data retrieval method introduces novel vulnerabilities since the information gathered by the retriever is largely unaudited and may contain poisoned data or data that can lead to unsafe responses from the LLMs.
        
\subsubsection{Model Tuning and Utilization}
\vspace{0.5cm}
After pre-training, the parameters of LLMs can be further updated by training on domain-specific datasets in downstream tasks. This process is known as fine-tuning (FT)~\cite{brown_language_2020}. A kind of fine-tuning method called supervised fine-tuning (SFT), aims to improve LLMs' responsiveness to instructions, ensuring more desirable reactions involving three major components of instructions, inputs, and outputs. Inputs relate to prompting and the inputs depend on the instructions, similar to applications of open-ended generation in ChatGPT. By providing both inputs and outputs they form an instance, and multiple instances can exist for a single instruction~\cite{he_large_2024}. Among fine-tuning, other training techniques within model prompting include instruction tuning and alignment tuning. By FT from a mixture of multi-task datasets formatted via natural language descriptions with the use of instruction tuning, LLMs are enabled to follow task instructions for new tasks without needing explicit examples, highlighting the ability of generalization for instruction following~\cite{zhao_survey_2023}. However, LLMs can demonstrate versatility, even without FT where they produce a phenomenon known as zero-shot learning, exhibiting the ability to perform tasks for which the model was never explicitly trained~\cite{brown_language_2020}.

Alignment tuning, equipped with reinforcement learning, is used to enhance LLMs to be safe interactive models. Since LLMs are trained to capture the data characteristics of uncurated pre-training corpora involving both high-quality and low-quality data, the LLM can generate toxic, biased, or harmful content for humans. To mitigate this problem, an FT process based on reinforcement learning from human feedback (RLHF) is used to align the LLM with the outcomes that satisfy human values~\cite{zhao_survey_2023}. The RLHF process ranks LLM outputs, with rewards scaled to positive and negative values. The LLM is then trained to produce highly-ranked responses and avoid low-ranked responses. In healthcare, RLHF provides advantages to the model such as improved accuracy and reliability through continuous feedback from medical professionals, and customizes the interactions based on real clinical settings and patient needs~\cite{he_survey_2024}. These advanced training techniques improve LLM's ability to generalize across tasks and improve their overall utility in various domains.

\begingroup
\begin{table*}[t]
    \caption{\textbf{Differences in Definitions of Safety.} There is no unifying definition of safety within the area of large language models. We see obvious agreement that models should be law-abiding, ethical, and non-violent in order to be safe, and as such these properties are strongly relevant to our definition of safety. However, beyond that point, there is generally a deviation between the authors' respective definitions. This creates two further categories of terms, properties that are moderately relevant to safety and those that are weakly relevant. Questions of fairness, the informing ability of an LLM, and robustness are generally covered but not unanimously, and hence are moderately relevant, whereas privacy-preserving properties or non-sycophancy are rarely discussed in the current literature and are thus weakly relevant to safety. This dialogue between different modes of thought concerning what makes a large language model ``safe" heavily influences our definition of safety and our resulting discussion.}
    \label{tab:def_safety_table}
    \centering
    \setlength{\tabcolsep}{1.75pt} 
    \renewcommand{\arraystretch}{1} 
    \resizebox{\textwidth}{!}{%
    \begin{tabular}{clccccccccc}
    \textbf{Relevance} &
     \textbf{Property} &
     \makecell{\textit{Sun et al.} \\ \cite{sun_trustllm_2024}} &
     \makecell{\textit{Liu et al.} \\ \cite{liu_trustworthy_2024}} &
     \makecell{\textit{Han et al.} \\ \cite{han_towards_2024}} &
     \makecell{\textit{Röttger et al.} \\ \cite{rottger_safetyprompts_2024}} &
     \makecell{\textit{Zhang et al.} \\ \cite{zhang_safetybench_2023}} &
     \makecell{\textit{Wang et al.} \\ \cite{wang_decodingtrust_2023}} &
     \makecell{\textit{Tedeschi et al.} \\ \cite{tedeschi_alert_2024}} &
     \makecell{\textit{Inan et al.} \\ \cite{inan_llama_2023}} &
     \makecell{\textit{Weidinger et al.} \\ \cite{weidinger_sociotechnical_2023}} \\
    \cmidrule{1-11}
     \multirow{3}{*}{Strong} &
    Ethical &
      \checkmark &
      \checkmark &
      \checkmark &
      \checkmark &
      \checkmark &
      \checkmark &
      \checkmark &
      \checkmark &
      \checkmark \\
    \cmidrule{2-11}
      &
    Law-abiding &
      \checkmark &
      \checkmark &
      \checkmark &
      \checkmark &
      \checkmark &
       &
      \checkmark &
      \checkmark &
      \checkmark \\
    \cmidrule{2-11}
      &
    Non-violent &
      \checkmark &
      \checkmark &
       &
      \checkmark &
      \checkmark &
      \checkmark &
      \checkmark &
      \checkmark &
      \checkmark \\
    \cmidrule{1-11}
     \multirow{3}{*}{Moderate} &
    Fair &
       &
       &
       &
      \checkmark &
      \checkmark &
      \checkmark &
      \checkmark &
       &
      \checkmark \\
    \cmidrule{2-11}
    &
    Informing &
       &
      \checkmark &
       &
       &
      \checkmark &
      \checkmark &
       &
       &
      \checkmark \\
    \cmidrule{2-11}
      \multirow{3}{*}{\vspace{-1cm} Weak} &
    Robust &
       &
       &
       &
      \checkmark &
      \checkmark &
      \checkmark &
       &
       &
      \checkmark \\
    \cmidrule{1-11}
     &
    Privacy Preserving &
      \checkmark &
       &
       &
      \checkmark &
       &
      \checkmark &
       &
       &
      \\
    \cmidrule{2-11}
       &
    Non-sycophantic &
       &
       &
       &
      \checkmark &
       &
       &
      \checkmark &
      \checkmark &
        \\
    \cmidrule{1-11}
    \end{tabular}
    }
\end{table*}
\endgroup

\section{Research Methodology}
\label{sec:methods}
\vspace{0.5cm}
The discussion of blockchain technology's incorporation into large language models necessitates a corresponding exploration into the implications of various terms and definitions found at that intersection. For example, due to the rapid emergence of LLMs, there exists an absence of consensus in describing common phenomena concerning LLM safety and security. To ameliorate this effect, we take care to stress opposing, but related, definitions of safety found within many different works in Table \ref{tab:def_safety_table}. In light of these distinctions, we offer two formal definitions of security and safety in order to contextualize these differing but similar areas of research. These definitions will also serve to highlight where particular blockchain technologies could be applied in their respective domains, and focus research efforts. 

First, and to allow a richer discussion centered around safety and security, we delineate between active and passive privacy within LLMs as introduced in~\cite{yan_protecting_2024}.
\begin{itemize}
\item \label{activeprivacy} \textit{Active privacy} is where a user intentionally tries to gain access to sensitive information by breaking the large language model, especially with backdoor attacks, prompt injection attacks, and membership inference attacks during the pre-training and FT phases.
\item \textit{Passive privacy} is the state or condition of any impacted person being protected from accidental or unexpected data leakage originating from a large language model. This definition includes protecting the privacy of not only users but people whose information was added to a model's corpus without their knowledge or consent.
\end{itemize}

Next, we introduce our definitions of security and safety regarding LLMs.
\begin{definition}
\label{security_def}
    \textbf{LLM Security.} A large language model is considered secure if it: 
    \begin{enumerate}
        \item Withstands applicable adversarial attacks and maintains system integrity, providing consistent and accurate responses, and 
        \item Ensures active user privacy, explicitly resisting backdoor, prompt injection, and inference attacks to prevent malicious users from extracting private information.
    \end{enumerate}
\end{definition}
\begin{definition}
\label{def_safety}
    \textbf{LLM Safety.} A large language model is considered safe if it interacts with users in a trustworthy manner, adhering to the aforementioned (Table~\ref{tab:def_terms_table} interrelated properties of safety: being ethical, law-abiding, nonviolent, fair, passively privacy-preserving, and informing.
\end{definition}
    
These definitions will serve a versatile role throughout this paper as building blocks for our contextualization of relevant and notable research efforts in BC4LLMs. Besides, they will serve the community at large in helping to establish reliable and tangible properties of secure and safe large language models. Furthermore, They will help establish tighter definitions of finer-grained terms and ideas within BC4LLMs. For example, in Table~\ref{tab:def_safety_table}, we provide definitions for terms found within our definition of safety to lessen the effect of the vague nature of some of the words. These definitions are backed by relevant examples found in the literature.

\begin{table*}[t]
    \caption{\textbf{Safety Area Definitions and Examples.} The area immediately surrounding BC4LLMs lacks a unifying definition of safety as well as consensus on what terms within that definition precisely mean. We provide generalized definitions for terms considered in our definition, as well as examples of incidents in literature where LLMs deviate from behavior as described in the definition. Italicized terms indicate inclusion in our definition of safety.}
    \label{tab:def_terms_table}
    \centering
    \scalebox{.9225}{
    \begin{tabular}{p{0.15\textwidth} p{0.325\textwidth} p{0.48\textwidth} l@{\,}l}
        \toprule
        \textbf{Safety Area} & \textbf{Definition} & \textbf{Example of Non-alignment} \\
        \cmidrule{1-3}
        \textit{Ethicality} & LLMs aligning with moral principles. & A LLM agreeing with eugenics.~\cite{hendrycks_aligning_2020}. & \rdelim\}{15}{1em} & \multirow{6}{*}{\rotatebox[origin=c]{90}{Found within definition of safety\hspace{.45cm}}} \\
        \cmidrule{1-3}
        \textit{Legality} & LLMs refusing to assist users in illegal endeavors. & A LLM assisting a user in creating incendiary devices.~\cite{tonkin_chatgpt_2023}. & & \\
        \cmidrule{1-3}
        \textit{Non-violence} & LLMs soliciting generally non-violent advice or instructions. & A LLM advising a user to perform a 'raid on a drug house' and 'kill everyone there'~\cite{ganguli_red_2022}. & & \\
        \cmidrule{1-3}
        \textit{Passive Privacy} & LLMs protecting private data within their corpus absent of malicious threats.& A LLM partially or fully reconstructing private images from a given dataset~\cite{carlini_extracting_2023}. & & \\
        \cmidrule{1-3}
        \textit{Honesty} & LLMs refraining from producing inaccurate or misinformed responses which may lead to negative outcomes. & LLMs administering faulty or fundamentally dangerous advice to patients or physicians in a healthcare setting~\cite{pal_med-halt_2023}. & & \\
        \cmidrule{1-3}
        \textit{Fairness} & LLMs ensuring a equitable environment for interaction, regardless of social identity. & LLMs associating ``male" names with qualities of leadership, and ``female" names with qualities of amicability~\cite{wan_kelly_2023}. & &\\
        \cmidrule{1-3}
        Robustness & The ability of the LLM to defend against adversarial attacks, originating from outside the model. This is a wide-reaching term, and falls within our discussion of security as it relates to LLMs. & A LLM falling victim to a backdoor attack planted in poisoned training data and producing malicious outputs as a result~\cite{yao_survey_2024}. \\
        \cmidrule{1-3}
        Non-sycophancy & LLMs choosing consistent outputs despite the chance that they may be in conflict with a user's beliefs or desires. & A LLM revising a correct answer to an incorrect answer after the user asks the LLM if they are sure or challenges the LLM's result in some way~\cite{sharma_towards_2023}. \\
        \bottomrule
    \end{tabular}%
    }
\end{table*}

\subsection{Research Approach and Limitations}
\vspace{0.5cm}
Literature surveys often are limited in their depth and scope by unconscious factors that impact the authors' ability to fairly select papers for review. To be transparent, and to aid researchers conducting similar or future reviews, we outline our research approach and its associated limitations. While conducting our research, we used the search engine, Google Scholar, and several databases, including ACM Computing Surveys, IEEE Xplore, SpringerLink, and arXiv. We chose these databases as they either produce quality research and contribute to the growth of interest in novel areas, or in the case of arXiv have the most up-to-date papers available. With Google Scholar, we used keyword searches such as ``blockchain for LLMs" and ``blockchain-based LLMs" as starting points for relevant, intriguing research papers. To solve problems concerning the scope and interrelated domains of disparate areas, we gathered various applications of blockchain for AI, blockchain-enabled machine learning, federated learning, and deep learning tactics to apply them to LLMs. Lastly, of note is the fact that we were largely aided in this further research effort by a waterfall approach to finding research papers. That is, we found several foundational papers in the BC4LLM field, explored citations in those papers, and subsequently explored citations in those secondary papers. We continued investigating relevant citations in this waterfall fashion until we reasonably exhausted all relevant articles. Admittedly, this method of finding prominent research articles is limited in its natural tendency to develop blind spots to less well-known research articles or venues. However, in the spirit of a literature survey, we choose to focus on more established papers that more accurately capture the trends currently found in the space. For our exclusion criteria, we limited our research as follows: no duplicates; found articles from 2016 and above, excluding the original Merkle Tree paper~\cite{merkle_digital_1988}; no Masters or Ph.D. theses; and only studies written in English.

\subsection{Model of Threat Categorization}
\begin{figure*}[h!]
    \center
    \includegraphics[scale = 0.55]{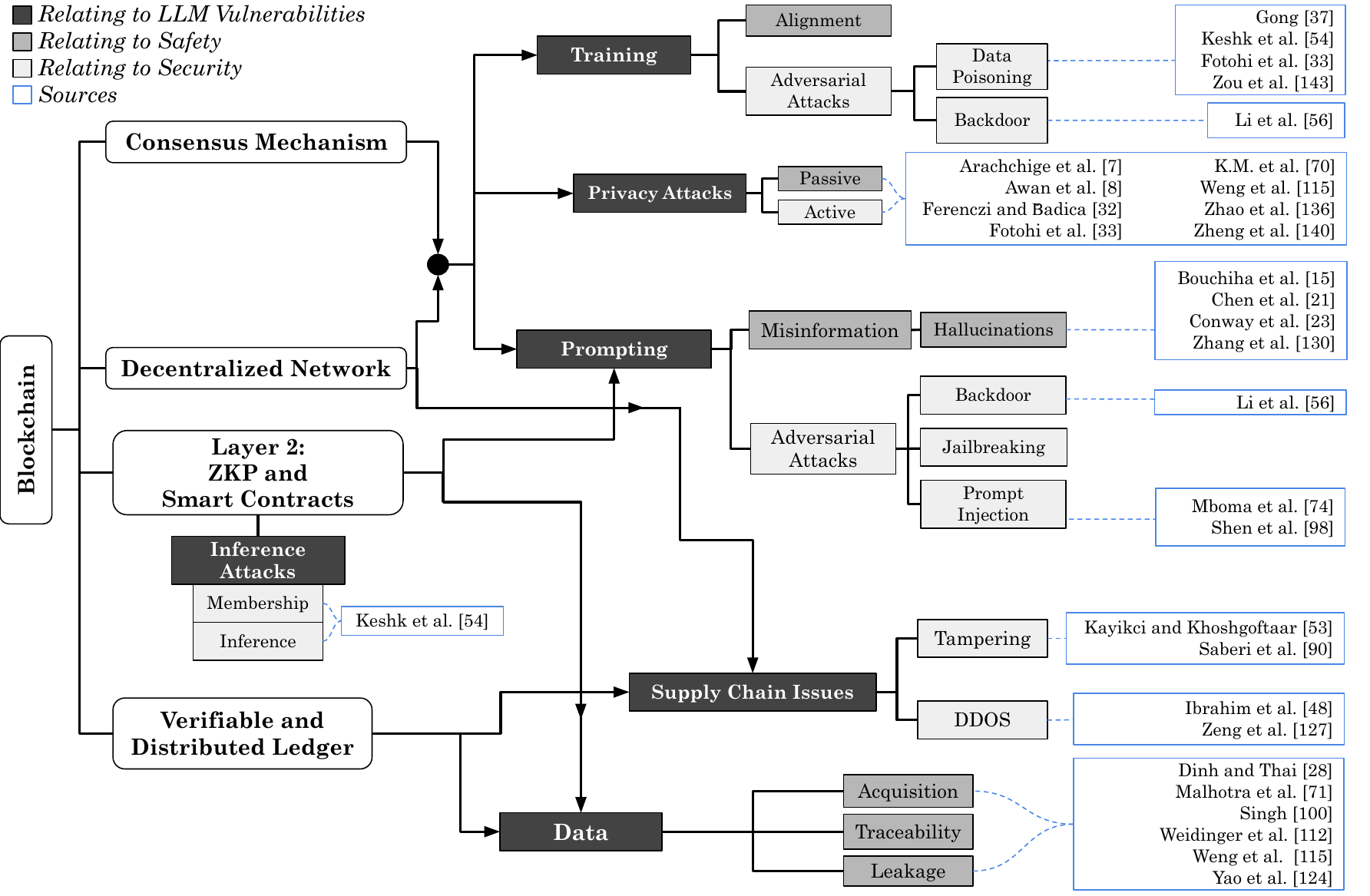}
    \caption{Taxonomy of Blockchain for LLM's Security and Safety. This diagram outlines the integration of blockchain technology to enhance the security and safety of large language models through categorizing interactions and safeguards into several layers and components. Each section supports relevant sources for further reference, illustrating a comprehensive approach to mitigating and preventing vulnerabilities as well as supplementing the security and safety of LLMs through blockchain technology. Promising areas that currently do not have blockchain as a solution to these vulnerabilities intentionally do not have source boxes.}
    \label{fig:Tax}
\end{figure*}

\vspace{0.5cm}        
There exists a wide variety of threats that affect LLMs. Oftentimes, many of these threats originate from the nature of LLMs acting as AI systems. In Figure~\ref{fig:Tax}, we refer to these vulnerabilities similarly to that discussed in~\cite{yao_survey_2024}, which categorized the most LLM vulnerabilities and AI-inherent vulnerabilities together, yet also included external threats under non-AI inherent vulnerabilities. We contribute further by applying these vulnerabilities to each respective process within developing a large language model and tying these to respective applications of blockchain. Beginning with the training process, LLMs are prone to threats such as data poisoning and backdoor attacks. As defined by Yao et al.~\cite{yao_survey_2024}, data poisoning is where attackers influence the training process by injecting malicious data into the training set, introducing vulnerabilities within the security and effectiveness of the model. Following the trend of poisoned data, there can be backdoor attacks implemented on the training data, as defined by Li et al.~\cite{li_blockchain-based_2023}, who categorize backdoor attacks into attacks on training data and attacks on local models. The backdoor attacks on training data are further divided into attacks based on label flipping and attacks based on planting triggers. Attacks based on label flipping focus on manipulating the labels, whereas attacks on planting triggers modify the input data and labels, effectively constructing an adversarial sample. Then, attacks on local models are further divided into attacks based on modifications to the training process and attacks based on manipulating the trained model~\cite{li_blockchain-based_2023}. The backdoor attacks can be applicable to both the training and prompting phases of LLMs when using this distinction.

\label{ThreatCategorization}

RAG attacks have a variety of issues, including privacy issues~\cite{zeng_good_2024, anderson_is_2024} and knowledge poisoning attacks~\cite{zou_poisonedrag_2024}. For RAG specifically, Xue et al.~\cite{xue_badrag_2024} propose BadRAG to identify security vulnerabilities, exposing direct attacks on the retrieval phase from semantic triggers, and uncovering indirect attacks on the generative phase of LLMs that were caused by a contaminated corpus. These RAG-specific attacks and defenses are elaborated on in Section~\ref{sec:BC_threats_LLMs}. When interacting with an LLM, AI's inherent vulnerabilities become evident, as highlighted in~\cite{yao_survey_2024}, since LLMs are fundamentally AI models themselves. We focus on the prevalent adversarial attacks that malicious users may use to tamper with the LLM, attempt to find out sensitive information, or ruin the system entirely. We recognize jailbreaking and prompt injection as two separate but similar types of adversarial attacks that are initiated within prompting. For instance, jailbreaking prompts are designed to bypass the restrictions set by service providers during model alignment or other containment approaches~\cite{shayegani_survey_2023}. Prompt injections aim to override an LLM's original prompt and direct it to follow a set of malicious instructions, leading to erroneous advice or unauthorized data leakage~\cite{liu_prompt_2024}. In Sections~\ref{sec:BC4PP} and~\ref{BC4Misinformation}, we discuss instances of misinformation and passive privacy leakage addressed as safety concerns. Note that we include backdoor attacks based on modifications to the trained model in prompting since these backdoor attacks can still happen after model training~\cite{li_blockchain-based_2023}.

Another relevant attack is a membership inference attack (MIA), a type of privacy attack where some malicious users, given access to the model, can determine whether a given point was used to train that model with high accuracy~\cite{neel_privacy_2024}. However, Neel and Chang~\cite{neel_privacy_2024} state that this attack is more related to information about the training point data leaking through the model, and that malicious users must have access to a candidate point in order to run the attack. Therefore, this attack is more prevalent with passive privacy, highlighting the need to prevent data leakage. Similar attacks are user inference attacks that seek to gain knowledge or insights about the model or data's characteristics, often by observing the model's responses or behavior~\cite{yao_survey_2024}.
    
Last but not least, we explore denial of service (DoS) attacks and supply chain vulnerabilities. Yao et al.~\cite{yao_survey_2024} describe DoS attacks as a type of cyber attack that aims to exhaust computational resources, resulting in latency or making the technology resources unavailable. In this survey, we focus on distributed denial of service attacks (DDoS), a type of DoS attack where requests flood the system, attacking simultaneously from multiple sources on the network~\cite{elmamy_survey_2020}. Yao et al.~\cite{yao_survey_2024} also defined LLM supply chain vulnerabilities as the risks in the lifecycle of LLM applications that may occur from using vulnerable components or services, including third-party plugins that may be used to steal chat histories, access private information, and or execute code on a user's machine. All of these security vulnerabilities are substantial threats to LLMs that need to be mitigated or prevented. Possible methods of defense are discussed in Section~\ref{sec:BC4LLMs_Security}, using current blockchain frameworks and experiments for these security problems, as listed by each developmental phase of the LLM, AI inherent threats, and supply chain issues. 
    
\section{Existing Literature on BC4LLM}
\label{sec:related_works_bc4llm}
\vspace{0.5cm}
Independently, the fields of both LLMs and blockchain research have grown substantially over the past several years. It is no surprise that the literature surrounding these topics has begun to morph and relate to each other. In previous research, we have seen LLMs for Blockchain Security~\cite{he_large_2024} as well as an introduction to the term BC4LLM in Luo et al~\cite{luo_bc4llm_2023} where they provide a comprehensive survey of blockchain for LLMs. However, they do not acknowledge the multitude of safety and security solutions that blockchain provides for certain LLM vulnerabilities. Effectively, Luo et al.~\cite{luo_bc4llm_2023} aim to introduce BC4LLM for trusted AI, enabling reliable learning corpora, secure training processes, and identifiable generated content. In juxtaposition, our survey aims to analyze possible BC4LLM solutions closely related to our definitions of safety (\ref{def_safety}) and security (\ref{security_def}) when looking at inherent system vulnerabilities in LLMs. To begin our analysis, we define these security problems based on previous work and highlight areas of research that are applicable to areas of BC4LLM safety and security. 

\subsection{Blockchain for LLMs' Security}
\label{sec:BC4LLMs_Security}
\vspace{0.5cm}
Few papers and experiments analyze how the integration of these two technological powerhouses interacts with one another. We have seen benefits of this integration that apply to our definition of security (\ref{security_def}). Balija et al.~\cite{balija_building_2024} introduce a peer-to-peer (P2P) federated LLM, namely PageRank, which works with a blockchain. This system operates in a fully decentralized capacity. Demonstrably, the blockchain implementation led to more efficient accuracy and latency results. With that being stated, Balija et al.~\cite{balija_building_2024} provide a developing direction in the field of BC4LLM to enhance system security. Below, we address several current vulnerabilities in LLMs and analyze them individually. In order to better understand these security problems, we categorize these vulnerabilities to their respective LLM training stages, highlight blockchain for AI works, and provide well-researched blockchain applications as a solution. 

Vulnerabilities are present at each step in the process of developing a LLM. In early methods of model training, we encounter adversarial attacks such as data poisoning and backdoor attacks within the corpus~\cite{shayegani_survey_2023, yan_protecting_2024}. Progressing into model fine-tuning and general use, the LLM can fall victim to prompt-based attacks~\cite{zhao_prompt_2023, abdali_securing_2024, yan_protecting_2024, liu_prompt_2024}, inference attacks~\cite{keshk_privacy-preserving-framework-based_2020, hu_membership_2022}, and RAG-related attacks~\cite{xue_badrag_2024, cheng_trojanrag_2024, anderson_is_2024, deng_pandora_2024, zou_poisonedrag_2024, zeng_good_2024}. These attacks are common vulnerabilities in both LLMs and AI since LLMs and AI are closely related as seen in Figure~\ref{fig:LLMandAI}. Some of the threats against LLMs can be addressed by implementations from blockchain for AI (BC4AI) research, as elaborated below in Section~\ref{sec:BC4AI}. Considering the volume of potential attacks against LLMs, we make a further distinction of solutions that are specifically related to BC4LLM research and other blockchain-based solutions from BC4AI research. With this, we are able to highlight shared vulnerabilities for LLMs and AI. We provide an analysis of how blockchain can help defend against and mitigate these vulnerabilities, starting with threats during each phase of LLM training and utilization, continuing onto different blockchain solutions for AI inherent threats, and lastly noteworthy technology inherent attacks such as denial of service (DDoS) attacks and issues with supply chain logistics. 
        
\subsubsection{Blockchain for Threats in LLMs' Training}
\label{BC4LLMTraining}
\vspace{0.5cm}
To mitigate the threats in LLMs' training, data selection stands as a crucial aspect of model development, with particular emphasis on ensuring that training data is authentic, safe, and resistant to data poisoning attacks.
One potential approach could be enabling the LLM to ``unlearn" poisoned data or data deemed unsafe based on our definition of safety~\ref{def_safety}. Zuo et al.~\cite{zuo_federatedtrustchain_2024} establish federated TrustChain to enhance LLMs' training and unlearning through a blockchain-based federated learning framework. Through integration with Hyperledger, the framework can efficiently perform unlearning, reducing the accuracy to 0.70\% after unlearning given that the initial accuracy is 99.15\%~\cite{zuo_federatedtrustchain_2024}. This demonstrates the potential of applying blockchain techniques to improve the security and privacy of LLMs, where LLMs can selectively forget specified data points while simultaneously preserving the performance via Low-Rank Adaptation (LoRA) and tuning hyper-parameters. This method of a blockchain-enabled federated unlearning process is further detailed as a future research possibility that has been thoroughly explored by few, as emphasized later in Section~\ref{BC4Unlearning}.
         
Another significant issue is data poisoning. To address this issue, Gong et al.~\cite{gong_dynamic_2023} propose a possible blockchain solution, introducing dynamic large language models (DLLM) on blockchains. Instead of using the traditional centralized datasets that LLMs are provided with, developing LLMs on blockchains enables the creation of decentralized datasets. These datasets are less likely to be tampered with and can be easily audited for accuracy. Gong et al.~\cite{gong_dynamic_2023} present the DLLM to evolve after the training process. 
This was implemented by adjusting neural network parameters, enabling the LLM to continue learning during its use.

\begin{figure}[h]
    \center
    \setlength{\belowcaptionskip}{-8pt}
    \includegraphics[scale = 0.95]{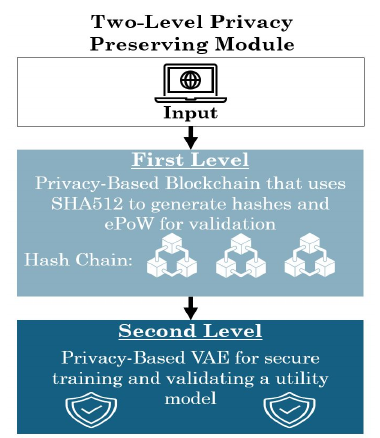}
    \caption{An overview of a two-level framework, consisting of a privacy-based blockchain that uses the secure hash algorithm 512 (SHA512) to generate hashes for data integrity. Then, the enhanced proof of work (ePoW) is used to authenticate data records and prevent data poisoning attacks from altering original data. These hashes of data blocks are linked to each other, called a Hash Chain. Then, for the second level, a privacy-based variational autoencoder (VAE) for secure data transformation ensures robust protection against inference attacks while maintaining the utility model for anomaly detection.}
    \label{fig:privacyframework}
\end{figure}
        
Additionally, blockchain-based systems can help assess where data poisoning may occur, and as shown in\cite{keshk_privacy-preserving-framework-based_2020}, blockchain can protect datasets and detect potential inference attacks through a two-level privacy preserving module. This research proposes a framework based on blockchain and deep learning, including two levels of privacy mechanisms as shown in Figure~\ref{fig:privacyframework}. For the first level, Keshk et al.~\cite{keshk_privacy-preserving-framework-based_2020} used SHA512 to generate secure hashes and then implemented an enhanced-proof-of-work (ePoW) technique for authenticating and preventing data poisoning attacks. The second level consisted of a VAE model for converting original data into an encoded format for mitigating inference attacks that could be learned from system-based machine learning. In their testing, these mechanisms were effective in preventing data poisoning and inference attacks from manipulated smart power network datasets. BC4LLMs could benefit from this similar type of implementation, working with secure methods of hashing and blockchain-based deep learning privacy preservation techniques. By integrating a two-level privacy-preserving module, BC4LLMs can ensure data integrity and confidentiality while effectively detecting and mitigating both data poisoning and inference attacks.

Poisoned data has been an interest with RAG in particular. For example, Xue et al.~\cite{xue_badrag_2024} develop a way to identify security vulnerabilities from a poisoned corpus, but they do not use blockchain as a solution. We address the absence of research on blockchain and RAG, especially when using blockchain to help prevent RAG security issues. We discuss this as a possible future research direction as there remains a current gap in research of blockchain-based RAG systems and elaborate on this topic in Section~\ref{BC4RAG}. Poisoned data overall is a major concern within LLMs and we offer blockchain as a potential source of ground truth to aid in mitigating this threat during the pre-training stages of LLMs and potentially mitigate RAG security concerns. In addition to data poisoning, LLMs are susceptible to backdoor attacks hidden in the training data during the LLM pre-training phase. Zhao et al.~\cite{zhao_prompt_2023} introduce ProAttack which improves the stealth of backdoor attacks by accurately labeling poisoned data samples. As these attacks improve and become more sophisticated, it is crucial to explore robust defense mechanisms for LLMs. Few defense mechanisms using blockchain techniques have been studied, while Li et al.~\cite{li_blockchain-based_2023} propose a blockchain-based federated-learning framework (DBFL) that withstands backdoor attacks in a blockchain environment by incorporating an RLR aggregation strategy into the aggregation algorithm of a user and the addition of gradient noise to limit the effectiveness of backdoor attacks. The robustness of FL against backdoor attacks is enhanced by using various blockchain functions, including digital signature verification and simulation of chain resynchronization~\cite{li_blockchain-based_2023}.

\subsubsection{Blockchain for Threats in LLMs' Prompting and Utilization}
\label{sec:BC_threats_LLMs}
\vspace{0.5cm}
LLMs are often further trained through techniques such as instruction tuning, alignment tuning, and fine-tuning, each of which may introduce specific vulnerabilities, such as prompt injection~\cite{mbula_mboma_assessing_2023, shen_large_2023} and backdoor attacks~\cite{li_blockchain-based_2023}. These adversarial attacks are included under the term active privacy~\ref{activeprivacy} where a malicious user attempts to gain unauthorized access. Blockchain technology, with its inherent transparency and immutability, holds the potential to mitigate and defend against these vulnerabilities. For instance, blockchain's ability to ensure data integrity and provide traceability can play a pivotal role in defending against prompt injection attacks. Mbula et al.~\cite{mbula_mboma_assessing_2023} produce an overview of LLMs for blockchain and note the capabilities of blockchain, especially the transparency and immutability to provide a reliable audit trail of transactions to track and investigate any suspicious activities. Although not specifically focused on prompt injection, this approach showcases how blockchain can supplement security by offering a clear and immutable record of interactions. Applying this to BC4LLMs can help prevent suspicious users from continuously interacting with an LLM, allowing for traceability to stop the user from entering malicious prompts. We recognize prompt injection is a critical vulnerability in LLMs and AI-related systems, yet as noted in the survey from Shen et al.~\cite{shen_large_2023} few blockchain defenses for prompt injection are present in the current field of research. 

Inference attacks are also a critical concern for LLMs and active privacy, as malicious users may attempt to extract sensitive data from the model, hence why the Taxonomy~\ref{fig:Tax} has inference attacks standalone. As discussed previously in Section~\ref{BC4LLMTraining}, Keshk et al.~\cite{keshk_privacy-preserving-framework-based_2020} apply Blockchain and DL techniques to preserve privacy and prevent inference attacks through a framework as depicted in Figure~\ref{fig:privacyframework}. For more inference attack applicable work, a survey~\cite{hu_membership_2022} thoroughly discusses membership inference attacks on ML and provides a group of defenses including differential privacy, regularization, confidence masking, and knowledge distillation. In other related works, there are instances of blockchain-based differential privacy methods~\cite{zhao_blockchain-based_2021, han_blockchain-based_2020, park_blockchain-based_2021}, but current research that uses blockchain-based differential privacy frameworks to prevent inference attacks is limited; It is worth noting that the theoretical foundation and potential synergies of this combination are promising. Another area with limited research is blockchain as a defense for jailbreaking attacks, which exploit the inherent capabilities of LLMs to bypass restrictions. There are multiple articles defending LLMs from jailbreaking attacks, yet little to none fully include blockchain to prevent jailbreaking. Hu et al.~\cite{hu_blockchain-based_2021} explores a blockchain defense mechanism for malware checking on operating systems, indicating a possible direction for future research in integrating blockchain to defend against jailbreaking in LLMs. As previously explained in Section~\ref{ThreatCategorization}, backdoor attacks after the model has been trained are based on modifications to the trained model. The key blockchain-based federated-learning framework from Li et al.~\cite{li_blockchain-based_2023} discussed in detail in Section~\ref{BC4LLMTraining} used a combination of a blockchain environment and an RLR aggregation strategy to defend against backdoor attacks. This framework effectively coordinated FL processes and maintained learning security and user privacy. When testing backdoor attacks caused by malicious participants, the accuracy of the model increased when using the RLR aggregation strategy~\cite{li_blockchain-based_2023}. Given these findings, the possibility of leveraging blockchain transparency and immutability presents a robust mechanism for improving LLM security against active privacy threats. However, comprehensive integration and empirical validation of blockchain-based defenses in LLMs remain imperative to advance the field of BC4LLMs. 
        
\subsubsection{AI-intrinsic Threats and Defenses}
\label{sec:BC4AI}
\vspace{0.5cm}        
AI intrinsic threats apply to LLMs due to the proximity of LLMs and AI, as shown in Figure~\ref{fig:LLMandAI}. Blockchain for AI (BC4AI) is an emerging technology, with blockchain-based solutions already being researched as a secure way to establish trust in the Internet of Things (IoT)~\cite{singh_convergence_2020, cuomo_how_2020}. Before BC4AI, some previous works refer to the integration as ``Onchain AI"~\cite{dinh_ai_2018, conway_opml_2024}. Research of BC4AI encapsulates other machine learning techniques, such as blockchain-based federated learning and blockchain for deep learning. Federated Learning is an addition to machine learning, as noted in Figure~\ref{fig:LLMandAI}, where federated learning uses a privacy-preserving and decentralized approach to centralized systems.

\begin{figure}[h!]
    \center
    \includegraphics[scale = 0.75]{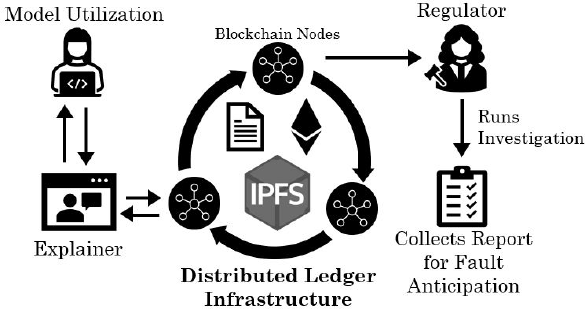}
    \setlength{\belowcaptionskip}{-10pt}
    \caption{Within BXAI, this diagram illustrates the framework that leverages a distributed ledger infrastructure, using the Ethereum Blockchain and IPFS for storage and secure, traceable transactions. Depicted is the interaction between model utilization: an explainer generating local post-hoc explanations, the storage of these explanations in IPFS, and their linkage to the Blockchain. The use of smart contracts helps secure and encrypt the data, then relay it to a regulator who investigates current explanations, ensuring accountability and fault anticipation. Blockchain nodes are used to facilitate the secure and transparent broadcast of events within the Ethereum network.}
    \label{fig:BXAI}
\end{figure}

A substantial literature on BC4AI has emerged from 2018 to 2024~\cite{dinh_ai_2018, wang_ai_2020, lopes_overview_2018, witt_blockchain_2024, fan_blockchain_2024, salah_blockchain_2019, taherdoost_blockchain_2022, malhotra_blockchain-based_2024, calvaresi_explainable_2019, baranwal_somy_ownership_2019, bhumichai_convergence_2024}. Among the most notable works, Salah et al.~\cite{salah_blockchain_2019} state the integration benefits of BC4AI. For example, there are five main benefits, such as enhanced data security, improved trust in robotic decisions, collective decision-making, decentralized intelligence, and high efficiency~\cite{salah_blockchain_2019}. For enhanced data security, information stored within a blockchain is considered highly secure. By storing sensitive and personal data in a distributed, disk-less environment, blockchain can work alongside AI algorithms to strengthen data protection and promote more trusted and credible decision outcomes. The other benefits of improving trust within AI decision-making involve using the blockchain as a record of the decision-making process, allowing better AI traceability to analyze the quality of responses. Secondly, Dinh and Thai~\cite{dinh_ai_2018} summarize the integration of blockchain and AI to where blockchain can assist AI in multiple aspects, as follows. AI can benefit in secure data sharing from blockchain, allowing transparency and accountability regarding which user's data is accessed, when, and by whom, letting users maintain control of their personal data. Among other data concerns, with the integration of blockchain and AI, blockchain technologies allow users to trade data via smart contracts, enabling the possibility of data marketplaces without a centralized middleman, making the transactions private and secure between users. Besides, Malhotra et al.~\cite{malhotra_blockchain-based_2024} propose a blockchain-based proof-of-authenticity framework for explainable AI (XAI) utilizing a public Ethereum Blockchain, smart contracts, and IPFS (Interplanetary File System) to ensure secure, traceable, auditable transactions within the Ethereum network. This framework highlights three major components, smart contracts, an Ethereum and IPFS interconnected network, and a regulator, as depicted in Figure~\ref{fig:BXAI}. Using smart contracts can enable continuous monitoring and tracing by all peers, in the case of any rule violations there are prompt rebound transactions to restore the system to an optimal state. To address the size limitation of storage on the blockchain, as further discussed in Section~\ref{sec:corpus_BC}, Malhotra et al.~\cite{malhotra_blockchain-based_2024} apply unique IPFS hashes stored on the Ethereum Blockchain to access larger-sized explanations that are stored off-chain in IPFS. These hashes are encrypted with the SHA256 algorithm to maintain data security. Thus, only entities with the corresponding hash can access and retrieve the IPFS hash and the associated explanation, ensuring controlled access even in a distributed network. Lastly, the regulator's role is responsible for auditing and has access to the explanations to predict the user at fault using audit trails if system failure were to occur.

\subsubsection{Non-AI Threats and Defenses}
\vspace{0.5cm}
Referring to Figure~\ref{fig:Tax}, we specifically focus on DDoS attacks and supply chain issues. Even though these attacks are common problems, we consider threats relevant to BC4LLMs. Ibrahim et al.~\cite{ibrahim_ddos_2022} suggest using a public blockchain to prevent DDoS attacks on IoT devices. Blockchain provides a tamper-proof platform as well as demonstrates how IoT devices working with blockchain can verify and authenticate using a trusted white-list which is implemented in the smart contract. Following this smart contract usage, if LLMs were to use a trusted white list for users then we can try to prevent these malicious users from trying to access the LLM in certain circumstances that are mutually agreed upon. Additionally proven by Shah et al.~\cite{shah_blockchain_2022}, blockchain-based solutions play a vital role in mitigating DDoS attacks.

A point of consideration is how DDoS attacks that target the blockchain to make the blockchain unavailable would require sufficient computer resources. The fully decentralized architecture of the blockchain and the consensus protocol for new blocks ensure that the blockchain can still operate meanwhile several blockchain nodes could be offline~\cite{zhang_securityandprivacyonblockchain_2020}. Incorporating this architecture into LLMs would help prevent DDoS attackers, as the larger the blockchain network is, then the harder it would be for a DDoS attack to be successful. Moreover, blockchain is known as a distributed, immutable, and verifiable ledger technology that ensures transparency and traceability~\cite{saberi_blockchain_2019}. By utilizing blockchain for LLMs, we can help mitigate these supply chain vulnerabilities. The decentralization of the network can maintain the integrity of the system at all points, aiding in mitigating the risk of a single point of failure, a common problem with centralized systems~\cite{saberi_blockchain_2019}. Blockchain is offered as a solution if the LLM were to accidentally crash, or was purposefully attacked by an attempt at overwhelming the system, then the LLM would still be intact since it is blockchain-based, removing the single point of failure entirely. However, it is important to note that blockchain solutions for LLMs depend on the availability of the underlying LLM infrastructure. If the LLM server is malfunctioning or shuts down, then these blockchain mechanisms may not be applicable, highlighting the need for a robust and resilient supply chain. To solidify the supply chain, blockchain offers secure transactional data in sectors including supply chain management, healthcare, and federated learning~\cite{zuo_federatedtrustchain_2024}. For better supply chain management and data traceability, Kayikci and Khosgoftaar~\cite{kayikci_blockchain_2024} address the potential intersection of blockchain and ML. ML can aid in analyzing data from multiple sources and identify potential supply chain issues such as delays or quality issues before they occur. By using blockchain to create a transparent record of all supply chain transactions there are improvements in security, openness, traceability, and productivity~\cite{kayikci_blockchain_2024}. While blockchain presents a promising solution for enhancing security and defending against adversarial threats to LLMs, ongoing research and development are necessary to address the evolving landscape of threats and vulnerabilities.

\subsection{Blockchain for LLMs' Safety}
\vspace{0.5cm}
The growing dominance of LLMs as search engines~\cite{reid_generative_2024}, code writers~\cite{zheng_survey_2023}, and in many other roles has introduced unique challenges related to their safety. For instance, LLMs who advise users to engage in dangerous activities such as eating glass~\cite{harrer_attention_2023} or which easily reveal personally identifying information~\cite{kim_propile_2024} may be unsafe for users to interact with even in the absence of external threats. In this section, we rely on our proposed definition of safety (Definition~\ref{def_safety}) to explore relevant literature that incorporates blockchain technology into the various solutions surrounding LLM safety. 
        
\subsubsection{Blockchain for Passive Privacy in LLMs}
\label{sec:BC4PP}
\vspace{0.5cm}
Despite its novelty, the concept of passive privacy is crucial for ensuring the safety of LLMs. Some models risk leaking sensitive information, potentially exposing private data like government-issued ID numbers and patient records~\cite{pan_privacy_2020}. The severity of these leaks underscores the need for effective solutions to advance LLMs responsibly. In this regard, blockchain’s guarantees of data sovereignty, obfuscation, and traceability offer practical passive privacy benefits that align well with the requirements of LLMs. In particular, we observe blockchain-based privacy preservation techniques which originate in varying proximity to LLMs as seen in BC4LLMs itself~\cite{ullah_privacy_2023, wellington_basedai_2024, singh_enhancing_2024}, blockchain-enabled deep learning~\cite{zhu_blockchain-based_2019, keshk_privacy-preserving-framework-based_2020, weng_deepchain_2021,  weng_auditable_2023, shafay_blockchain_2023}, blockchain-enabled machine learning~\cite{arachchige_trustworthy_2020}, and blockchain-enabled federated learning~\cite{awan_poster_2019, nagar_privacy-preserving_2019, zhao_privacy-preserving_2021, qammar_securing_2023, ferenczi_fully_2023, qu_blockchain-enabled_2023, m_privacy-preserving_2024}.

Within our focus on BC4LLMs, we have observed distinct trends in the application of blockchain to LLMs in their capacity to bolster passive privacy guarantees. Most notably, the development of zero-knowledge LLMs, a.k.a. ZK-LLMs, as described in~\cite{wellington_basedai_2024} and~\cite{singh_enhancing_2024}, has the potential to drastically reduce privacy leakage risks when interacting with LLMs. Considering the problem of data leakage approached from the lens of access, this application is natural. A user querying for their own personally identifiable information should, ideally, be able to access it whereas an unauthorized user should not. Obfuscating portions or the entirety of a corpus using zero-knowledge proofs~\cite{zhang2020transparent} allows for untrusted training nodes, or the model itself, to act on sensitive data without the ability to regurgitate it to a potentially malicious party. This same mechanism has broad applications that have been explored in other recent works as well, with special focuses on ZKPs for data curation and pre-processing~\cite{ullah_privacy_2023}, which consequently enhance passive privacy within LLMs.

Additionally, besides material on BC4LLMs, it is necessary to discuss passive privacy contributions made within LLM-related areas, as described in our classification of LLMs in the context of AI, ML, and DL (Figure~\ref{fig:LLMandAI}). Especially important in its immediate applications to LLMs, blockchain-enabled deep learning (BC-DL) is a growing field with potentially large impacts on LLM's passive privacy. Specifically, certain BC-DL technologies propose learning mechanisms distinct from traditional federated learning models~\cite{keshk_privacy-preserving-framework-based_2020, weng_deepchain_2021, shafay_blockchain_2023}. The concerted research effort to develop efficient distributed learning models that deviate from the typical model of federated learning is clearly well underway. This field has broad implications for blockchain; through the utilization of various blockchain properties, we see the development of privacy guarantees which undoubtedly strengthens the BC4LLMs area.

\begin{figure}[h]
    \centering
    \includegraphics[width = 1\linewidth]{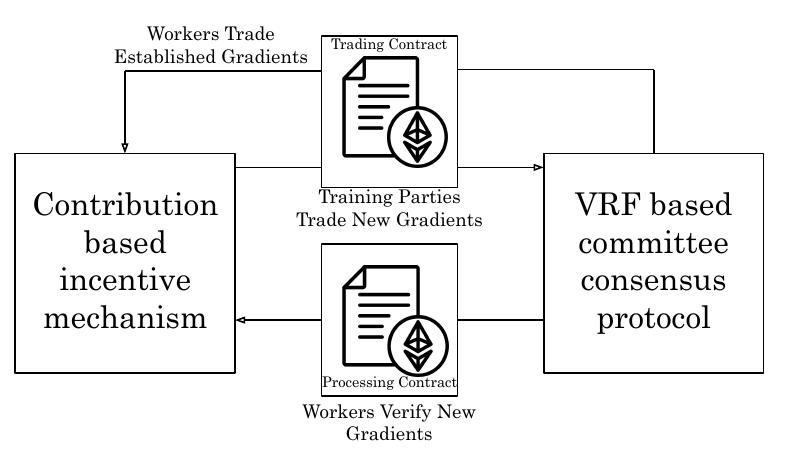}
    \caption{DeepChain deviates from conventional federated learning models by incorporating a synchronous requirement for consensus on data incorporated into the model during the final training round. The contribution-based incentive mechanism rewards participants for verifying new gradients and activates a trading contract to facilitate the sharing of updated gradient information. Besides, a Verifiable Random Function (VRF) ensures fairness in the committee-based consensus process, addressing concerns related to finality.}
    \label{fig:DeepChain}
\end{figure}

A noteworthy contribution in the field of blockchain and deep learning is the influential DeepChain~\cite{weng_deepchain_2021}, which introduces a novel privacy-preserving training framework based on blockchain technology. 
This system employs a consensus protocol alongside an incentive mechanism, enabling the use of private training gradients and ensuring the auditability of training data. This dual approach, incorporating zero-knowledge proofs in various aspects of the protocol, represents a promising new direction for blockchain-enabled passive privacy, building upon the well-established domain of federated learning. To underscore the novelty of this approach, Figure~\ref{fig:DeepChain} illustrates how the system operates primarily through consensus and incentivization mechanisms.

Additionally, despite this potential research direction, discussion on the wide body of research that does exist concerning blockchain-enabled federated learning (BC-FL) is essential when describing blockchain as a vehicle for improving the safety and security of LLMs. Federated learning, introduced by McMahan et al.~\cite{mcmahan_communication-efficient_2017}, has since been the principal building block of decentralized learning approaches in machine learning systems. While too broad to be considered for BC4LLMs, notable to this paper are the contributions of authors approaching BC-FL in its capacity as a powerful privacy preserving mechanism~\cite{nagar_privacy-preserving_2019, awan_poster_2019, zhao_privacy-preserving_2021, ferenczi_fully_2023, fotohi_decentralized_2024, m_privacy-preserving_2024}.

\subsubsection{Blockchain for Misinformation in LLMs}
\label{BC4Misinformation}
\vspace{0.5cm}        
The issue of LLMs fabricating information, commonly known as hallucinations~\cite{maynez_faithfulness_2020, openai_gpt-4_2024} is well understood. As a result, detecting and defending against hallucinations is a widely explored area~\cite{seneviratne_blockchain_2022}, with more research still yet to be conducted~\cite{ye_cognitive_2023, zhang_sirens_2023}. Along with this pressing general body of research, efforts have been made to leverage blockchain technology to reduce hallucinations by consensus-oriented~\cite{zhang_proof_2024} and oracle-based~\cite{bouchiha_llmchain_2024} approaches. Within consensus, Zhang et al.~\cite{zhang_proof_2024} proposed a system for efficient large language model inference quality assessment. That is, the veracity of a given model's responses was able to be assessed by using a  `Proof of Quality' consensus mechanism with low latency between the user and language model. This stands in contrast to other approaches, such as Bouchiha et al.'s~\cite{bouchiha_llmchain_2024} reputation-based system LLMChain, which relies on a decentralized oracle to cross reference request/response pairs originating from differing models and speak to the quality of inferences based on those comparisons. It is worth noting that despite these fundamental differences, they are both consensus-based approaches. This serves as an excellent example of how BC4LLM technology can take many different forms towards the same goal. 

In addition to advancing consensus-driven governance models for mitigating misinformation, several approaches have been explored to improve the accuracy of LLMs' responses. These efforts include contributions from the zero-knowledge domain, as demonstrated by Chen et al.~\cite{chen_zkml_2024}, more optimistic privacy assurances found in works like~\cite{conway_opml_2024}, and straightforward applications of verifiable ledgers, as proposed by Yazdinejad et al.~\cite{yazdinejad_making_2020}. Chen et al.~\cite{chen_zkml_2024} propose zkML, a compiler that enables TensorFlow models to be translated readily into zk-SNARK halo2 circuits via either KZG or IPA commitments. This conversion allows for any portion, or the entirety of, an LLM to gain the properties of zero-knowledge, knowledge soundness, and completeness. Through this, and with potential connections to verifiable databases, zkML gains the powerful ability to audit inferences and ensure their accuracy. This research avenue is particularly promising due to both the efficient and potentially on-chain verification of zk-SNARKs as well as the extensibility of zkML to virtually any ML model.

Distinct from the zkML approach is opML~\cite{conway_opml_2024}, which opts for an optimistic approach reliant on a fraud-proof rather than a ZK proof to catch erroneous outputs within a certain challenge period. Clearly, there exist trade-offs in this implementation when compared to the zkML approach. Optimistic rollups are desirable in the sense that they are performant, but if implemented in a RAG environment, or similarly situated between the user and a model, latency issues can quickly become dominant. Apart from proof-oriented mechanisms and worth noting is the work proposed by Yazdinejad et al.~\cite{yazdinejad_making_2020}, which focuses on detecting deepfakes using blockchain's verifiable ledger. While not directly applicable to the realm of LLMs due to the non-atomic nature of data within a language model, important insights can be drawn from the paper. Namely, BC4LLMs could benefit greatly from a proposed hashing method applied to particularly sensitive data areas such as names, addresses, or even health-care-related parts of corpora. This hash could be used as a guarantee of data veracity and could potentially prevent unsafe behaviors such as sycophancy, deception, or unfairness. Indeed, this hashing mechanism has the potential to be used as a final check for the LLM to verify that it is submitting information to the user that is consistent with standards agreed upon when information was originally committed to the ledger. Many similar vehicles for the maintenance of data integrity exist, albeit currently limited by scaling issues on-chain~\cite{deshpande_distributed_nodate}.

\section{Datasets Relevant to BC4LLM}
\label{sec:datasets}
\vspace{0.5cm}
Developing synergistic techniques that integrate blockchain with LLMs is essential for ensuring the safety and trustworthiness of future LLMs. In this context, it is critical to access to relevant datasets for experimentation. Blockchain-enabled systems often require unconventional training sets and edge cases to capture the dynamism and robustness of these implementations. Accordingly, we have compiled and summarized the relevance of specific datasets in Table~\ref{tbl:datasets}. While we include standard datasets such as MNIST, CIFAR-10, SQuAD, and MS-MARCO, we also highlight lesser-known datasets that may prove valuable in particular research contexts.

\vspace{.35cm}
\begin{table*}
\caption{\textbf{Datasets Relevant to BC4LLMs}}
\label{tbl:datasets}
\vspace{-.35cm}
\begin{tabular}{p{0.17\linewidth} p{0.10\linewidth} p{0.45\linewidth} p{0.20\linewidth}}
    \toprule
    \textbf{Dataset} & \textbf{Use Case} & \textbf{Description} & \textbf{Papers} \\
    \midrule
    MNIST\footnotemark[1] & \multirow{4}{=}{\vspace{1.75cm}\vfill \centering Pattern Recognition} & Images of handwritten digits for pattern recognition applications or vulnerability analysis. \vspace{0.1cm} & \cite{weng_deepchain_2021, weng_mystique_2021, mcmahan_communication-efficient_2017, ferenczi_fully_2023, shen_exploiting_2021, li_byzantine_2021} \\
    CIFAR-10\footnotemark[2] & & Labeled images used in capacities from improving pattern recognition to zk-SNARK benchmarks. \vspace{0.1cm} & \cite{weng_mystique_2021, chen_zkml_2024, mcmahan_communication-efficient_2017} \\
    MS MARCO\footnotemark[3] & & Collection of human answered questions, used in training corpora as well as simulating RAG attacks. \vspace{0.1cm} & \cite{xue_badrag_2024, zou_poisonedrag_2024, cheng_trojanrag_2024} \\
    MedMINST\footnotemark[4] & & Collection of medical images from case studies. \vspace{0.1cm} & \cite{xue_badrag_2024, zou_poisonedrag_2024, cheng_trojanrag_2024} \\ 
    \midrule
    Natural Questions\footnotemark[5] & \multirow{2}{=}{\vspace{.5cm}\vfill \centering Poisoned RAG} & Open domain question answering dataset, incorporating questions from users and rigorous answers. \vspace{0.1cm} & \cite{xue_badrag_2024, zou_poisonedrag_2024, cheng_trojanrag_2024} \\ 
    HotpotQA\footnotemark[6] & & Question answering dataset with multi-hop questions and supervised, regulated, answers. \vspace{0.1cm} & \cite{zou_poisonedrag_2024, cheng_trojanrag_2024} \\
    \midrule
    MT BENCH\footnotemark[7] & \multirow{2}{=}{\vspace{1pt}\vfill \centering LLM Evaluation} & Ranked pairwise expert human preferences for various model responses. \vspace{0.1cm} & \cite{zheng_judging_2023, bouchiha_llmchain_2024} \\
    SQuAD\footnotemark[8] & & Reading comprehension dataset comprised of questions posed on Wikipedia article with answers as sections of those corresponding articles. \vspace{0.1cm} & \cite{zuo_federatedtrustchain_2024, hu_prompt_2024} \\
    \midrule
    IMDB Dataset\footnotemark[9] & {\centering Sentiment Analysis\par} & Movie reviews \vspace{0.1cm} & \cite{zuo_federatedtrustchain_2024, hu_prompt_2024} \\
    \midrule
    SafetyBench\footnotemark[10] & {\centering Safety Evaluation\par} & Large number of multiple choice questions focused on evaluating the safety of LLMs. \vspace{0.1cm} & \cite{zhang_safetybench_2023} \\
    \midrule
    Tweets2011\footnotemark[11] & \multirow{8}{=}{\vspace{2.5cm}\vfill \centering Sensitive Information Handling} & List of scraped tweet identifiers and corresponding tweets from early 2011. \vspace{0.1cm} & \cite{zuo_federatedtrustchain_2024} \\
    MTSamples Scrape\footnotemark[12] & & Sample transcription medical reports from various disciplines and areas. \vspace{0.1cm} & \cite{zuo_federatedtrustchain_2024, hu_prompt_2024} \\
    DRC Diplomas\footnotemark[13] & & Highschool diplomas from the Democratic Republic of the Congo. \vspace{0.1cm} & \cite{balija_building_2024} \\
    HealthCareMagic\footnotemark[14] & & Real patient-doctor conversations found through the HealthCareMagic website, capturing the nature of patient vocabulary. \vspace{0.25cm} & \cite{anderson_is_2024} \\
    Enron Emails\footnotemark[15] & & Large set of emails generated by employees of the Enron Corporation. \vspace{0.1cm} & \cite{zuo_federatedtrustchain_2024, hu_prompt_2024} \\
    LLMGooAQ\footnotemark[16] & & Comprehensive database capturing question and answers from a wide variety of domains. \vspace{0.1cm} & \cite{bouchiha_llmchain_2024} \\ 
    GooAQ\footnotemark[17] & & Large scale question answering dataset aimed at developing a vast selection of question types. \vspace{0.1cm} & \cite{bouchiha_llmchain_2024} \\
    The Pile\footnotemark[18] & & Massive and open source data set consisting of a combination of roughly 20 other datasets. \vspace{0.1cm} & \cite{zhao_prompt_2023} \\ 
    \bottomrule
    \multicolumn{4}{l}{\footnotemark[1]https://yann.lecun.com/exdb/mnist/ 
    \footnotemark[2]https://www.cs.toronto.edu/~kriz/cifar.html 
    \footnotemark[3]https://microsoft.github.io/msmarco/} \\
    \multicolumn{4}{l}{\footnotemark[4]https://medmnist.com/ 
    \footnotemark[5]https://ai.google.com/research/NaturalQuestions 
    \footnotemark[6]https://hotpotqa.github.io/} \\
    \multicolumn{4}{l}{\footnotemark[7]https://paperswithcode.com/dataset/mt-bench 
    \footnotemark[8]https://rajpurkar.github.io/SQuAD-explorer/} \\
    \multicolumn{4}{l}{\footnotemark[9]https://developer.imdb.com/non-commercial-datasets/ \footnotemark[10]https://github.com/thu-coai/SafetyBench} \\
    \multicolumn{4}{l}{\footnotemark[11]https://trec.nist.gov/data/tweets/ \footnotemark[12]https://mtsamples.com/ \footnotemark[13]https://minepst.gouv.cd/palmares-exetat/} \\
    \multicolumn{4}{l}{\footnotemark[14]https://huggingface.co/datasets/RafaelMPereira/HealthCareMagic-100k-Chat-Format-en} \\
    \multicolumn{4}{l}{\footnotemark[15]https://huggingface.co/datasets/preference-agents/enron-cleaned \footnotemark[16]https://github.com/mohaminemed/LLMGooAQ/} \\ 
    \multicolumn{4}{l}{\footnotemark[17]https://huggingface.co/datasets/allenai/gooaq \footnotemark[18]https://pile.eleuther.ai/} 
    \end{tabular}
\end{table*}

\section{Challenges in BC4LLM}
\label{sec:challenges}
\vspace{0.5cm}
Despite the promise of the emerging BC4LLMs field, there are several innate challenges that delay progress and inhibit potential research directions. Typically, these are derived from certain limitations in blockchain technology, LLMs, or deficits in the way that blockchain can serve LLMs. 

\subsection{Corpus on Blockchain}
\label{sec:corpus_BC} 
\vspace{0.5cm}
LLMs' training heavily relies on substantial data, with modern corpora typically exceeding dozens of terabytes in volume~\cite{zhao_survey_2023}. This characteristic is inherently misaligned with the constraints that blockchain systems are generally designed to address. Reconciling blockchain's limitations in throughput and data-handling capacity with the extensive data requirements of LLMs represents one of the most pressing challenges in BC4LLMs. Several approaches have explored the use of zero-knowledge proofs (ZKPs) to enhance scalability~\cite{wellington_basedai_2024, singh_enhancing_2024}. However, relying on zero-knowledge technology solely for scalability, and not privacy, poses challenges due to the computational cost of generating ZKPs, even with minimal circuits. A significant factor here is the ongoing issue of the Multi-Scalar Multiplication (MSM) in ZKP generation~\cite{xavier_pipemsm_2022}. Furthermore, current WebGPU and WASM implementations likely fall short of the throughput required by client-based LLMs. For these reasons, it is improbable that zero-knowledge could serve as a definitive solution to scalability in BC4LLMs without significant advancements in zk-SNARK generation research. Addressing the discrepancy between the growing size of LLMs and blockchain’s limited capacity for on-chain data storage remains a substantial research challenge.
    
\subsection{Reliance on Oracles}
\vspace{0.5cm}
The security guarantees of blockchain technology, while robust, often leave little room for interoperability with external systems~\cite{deshpande_distributed_nodate}. That is, the blockchain can most easily interact with information on the chain, leaving little room to consider issues such as fact-checking or moral alignment. Oftentimes, to develop mechanisms that seek to assist with LLM toxicity or factuality, oracles are used to bridge this gap~\cite{fraga-lamas_fake_2020, fan_blockchain_2024}. Serving as mediators between chains and online sources, oracles are trusted parties that deliver information through a variety of protocols and frameworks. However, introducing a trusted party into an otherwise trustless system has been a long-standing weak point in this solution~\cite{mbula_mboma_assessing_2023}. Exploring non-oracle-based options for ground truth solutions, or toxicity checks, would greatly enhance the security guarantees of blockchain within LLMs. 

\subsection{Energy Consumption}
\vspace{0.5cm}
A significant portion of the challenges associated with LLMs arises from their need to consume and process vast amounts of data~\cite{zhao_survey_2023}. This requirement, in turn, necessitates extensive energy consumption during both training and runtime~\cite{luo_bc4llm_2023}. On the other hand, blockchain systems face their own energy challenges, as consensus mechanisms and transaction validation processes often incur substantial computational costs~\cite{mbula_mboma_assessing_2023}. The high computational demands of both LLMs and blockchain systems highlight a misalignment with the scalability of BC4LLM implementations without substantial efforts to reduce energy costs. This might require moving away from transformer-based architectures and energy-intensive consensus mechanisms, such as proof of work, toward more sustainable alternatives~\cite{nguyen_proof-of-stake_2019}.
    
\section{Future Research Directions}
\label{sec:future_directions}
\vspace{0.5cm}
There exist several critically overlooked areas within LLMs that may benefit greatly from the introduction of blockchain technologies. The most prominent of these areas include blockchain federated unlearning, RAG, differential privacy, data provenance, and toxicity mitigation.

\subsection{Blockchain Federated Unlearning}
\label{BC4Unlearning}
\vspace{0.5cm}
Privacy regulations are paramount in the online realm, especially concerning the ``right to be forgotten" and user data privacy which are critical considerations when working with LLMs and blockchain. Federated blockchain unlearning offers LLMs the ability to erase learned data. Within our research, we identified four recent papers that have implemented blockchain-based federated unlearning frameworks. As noted previously in Section~\ref{BC4LLMTraining}, Zuo et al.~\cite{zuo_federatedtrustchain_2024} develop a federated TrustChain framework for blockchain-enhanced LLM training and unlearning, focusing on the impact of Low-Rank Adaptation (LoRA) hyperparameters on unlearning performances and integrating Hyperledger Fabric to ensure the security, transparency, and verifiability of the unlearning process. In another study, Zuo et al.~\cite{zuo_federated_2024} presented a trustworthy approach towards federated learning with blockchain-enhanced machine unlearning. This implementation differs from Trustchain, where Zuo et al.~\cite{zuo_federated_2024} used a machine unlearning mechanism that utilized two types of clients for training and unlearning, smart contracts for process automation, and a blockchain network for secure, immutable record-keeping. Beyond the above works~\cite{zuo_federated_2024}, Liu et al.~\cite{liu_decentralized_2024} introduced Blockchain Federated Unlearning (BlockFUL) as a versatile framework that redesigns the blockchain structure using a Chameleon Hash (CH) technology to simplify model updates and reduce the computational and consensus costs associated with unlearning tasks. Additionally, BlockFUL ensures the integrity and traceability of model updates, including privacy-preserving results from these blockchain-based unlearning operations~\cite{liu_decentralized_2024}. Lin et al.~\cite{lin_blockchain-enabled_2024} propose a framework with a proof of federated unlearning protocol that also utilizes the Chameleon hash function to verify data removal and eliminate the data contributions stored in other clients' models. Both use CH functions in their blockchain-enabled federated unlearning processes. The applications of key blockchain components, such as on-chain smart contracts and hash mappings for verifying data removal, may enable LLMs to forget personal data effectively. Blockchain for unlearning is an emerging area of research with significant potential for further innovation.
    
\subsection{Blockchain-enhanced RAG}
\label{BC4RAG}
\vspace{0.5cm}
Considering the popularity of Retrieval Augmented Generation (RAG), extensive studies have emerged to focus on potential vulnerabilities within RAG that could compromise the integrity of LLMs~\cite{xue_badrag_2024, anderson_is_2024, deng_pandora_2024, zou_poisonedrag_2024, hu_prompt_2024, zeng_good_2024, cheng_trojanrag_2024}. However, a significant gap remains in addressing strategies to mitigate these attacks, particularly where blockchain technology could offer defensive benefits. Recent studies have called for exploring blockchain’s role in RAG deployment~\cite{balakrishnan_enhancing_2024}, and preliminary investigations have assessed blockchain’s potential to enhance user experience~\cite{zhang_interactive_2024} and performance evaluation~\cite{park_blockchain-based_2021}. Nonetheless, dedicated efforts to strengthen security and safety within RAG systems are largely absent in current literature. Advancing BC4LLMs specifically in the context of RAG security could yield considerable mutual benefits for both blockchain and LLM technologies.

\subsection{Blockchain for Privacy Guarantees in LLMs}
\vspace{0.5cm}
The clear connection between federated learning, blockchain, and LLMs allows for the field of differential privacy to enter BC4LLMs' sphere of relevance. Major contributions concerning the impact of differential privacy on related areas such as deep learning have already been made~\cite{abadi_deep-learning-dp}, but issues such as privacy budget exhaustion still loom large in the space~\cite{bkakria_budget-dp}. Moreover, despite conclusions that blockchain can help with privacy budget exhaustion~\cite{han_blockchain-based_2020, zhao_blockchain-based_2021}, few efforts have been conducted in exploring these solutions. Indeed, there is a need for more relevant research in order to realize the full measure of blockchain's impact on this area.

\subsection{Blockchain for Data Provenance and Transparency in LLMs}
\vspace{0.5cm}
Several recent papers have urged for increased data accountability measures to be placed on organizations developing LLMs, especially where it concerns issues of data acquisition~\cite{bender_dangers_2021, he_survey_2024}. Additionally, worth noting are direct calls for the introduction of blockchain technology to help solve the issue of data provenance~\cite{werder_establishing_2022} in LLMs. Largely, while this has been answered with responses in the realms of auditability~\cite{li_blockchain-based_2023}, straightforward data tracking solutions have remained absent from the literature, despite relatively simple conceptual formulations~\cite{dinh_ai_2018}. Towards this goal of achieving improved data provenance within LLMs corpus', RAG databases, and even in-context learning repositories, there is a need for more explorations into this natural application of distributed ledger technology to problems of explainable AI concerning LLMs.

\subsection{Blockchain for Non-toxic LLMs}
\vspace{0.5cm}
Encompassing vital attributes such as ethics, legality, and non-violence, developing non-toxic LLMs has been and will continue to be a major focus of the field for the foreseeable future~\cite{shen_large_2023, liu_trustworthy_2024, han_towards_2024}. There is no doubt that automated filtering of generated toxic content is one of the most pressing challenges concerning the safety of LLMs~\cite{gehman_realtoxicityprompts_2020, anwar_foundational_2024}. This is because in essence, filtering inferences negatively impacts the quality of LLM responses, whereas manual human annotation is a costly and complex process~\cite{anwar_foundational_2024}. Therefore, the applications of blockchain technology in this regard, while currently limited, are compelling. Considering one of the most groundbreaking achievements in ML within the past several years, federated learning has allowed for massive strides to be made within the spaces of securing training sets, user privacy, and even misinformation defense. A similar approach, aimed at the problem of toxicity, could be a hugely beneficial endeavor to the field. Moreover, imagining such a model is not difficult. Developing consensus around what is considered correct in a model and using that to propagate gradients and parameters is not dissimilar to the decisions that must be made about what is or is not toxic given the state of certain corpora. Given a concentrated research program, automated non-toxicity could very well have excellent solutions found in the blockchain space.

\section{Conclusion}
\label{sec:conclusion}
\vspace{0.5cm}
In this survey, we first highlight significant systemic vulnerabilities in large language models (LLMs), including data poisoning, hallucinations, jailbreaking, and privacy attacks. Although these issues have been extensively studied in conventional machine learning models, with approaches like differential privacy and federated learning, comprehensive protection for LLMs remains an area for improvement. In contrast, blockchain technology offers a promising solution to enhance the security and safety of LLMs. Blockchain systems provide powerful mechanisms to ensure data integrity, provenance, and encrypted frameworks, which can be leveraged to strengthen LLM defenses. By integrating blockchain-based defenses, it is possible to achieve stronger privacy protection, reliable data, and improved resilience of LLMs against adversarial threats.

Besides, it is critical to establish clear definitions of security and safety in the context of LLMs. We conclude that security for LLMs pertains to the ability to tolerate applicable adversarial attacks while maintaining system integrity to provide consistent and accurate responses, whereas safety for LLMs is the model's capacity to interact with users in a trustworthy manner, contingent upon adhering to ethical concerns, law-abiding, non-violent, fair, passively privacy-preserving, and informing. Additionally, differentiating between active and passive privacy measures will aid in developing more targeted and effective privacy-preserving strategies. These distinctions and definitions provide a foundational framework for future research in BC4LLM. From analyzing the integration of blockchain and LLMs, we propose a new taxonomy in Figure~\ref{fig:Tax}, where previous research done in the field of BC4LLMs can apply to security and safety problems that LLMs face. We recognize various gaps in BC4LLMs that need to be looked into for further consideration. In refining our understanding of relevant concepts, we see that the intersection of blockchain and LLMs holds significant potential for addressing the current shortcomings in LLM security and safety. Through our review, we aim to guide new researchers in understanding how blockchain technology can be utilized to enhance the security, reliability, and safety of LLMs.

\printbibliography

\end{document}